\documentclass[aps,prc,twocolumn,amsfonts,showpacs,floats,preprintnumbers,byrevtex,nofootinbib]{revtex4} 
\usepackage{epsfig}
\usepackage{dcolumn}
\newcommand{\nuc}[2]{$^{#1}${#2}}
\newcommand{\bd}{{\beta_2}}
\newcommand{\Qzt}{{Q_{c2}^{(t)}(J,k)}}
\newcommand{\Qzs}{{Q_{c2}^{(s)}(J,k)}}

\newcommand{\be}{\begin{equation}}
\newcommand{\ee}{\end{equation}}

\begin{document}
\title{Microscopic study of $^{240}$Pu, mean-field and beyond}
\author{M. Bender}
\affiliation{Service de Physique Nucl\'eaire Th\'eorique, 
             Universit\'e Libre de Bruxelles, C.P. 229, B-1050 Bruxelles, 
             Belgium}
\author{P.-H. Heenen}
\affiliation{Service de Physique Nucl\'eaire Th\'eorique, 
             Universit\'e Libre de Bruxelles, C.P. 229, B-1050 Bruxelles, 
             Belgium}
\author{P. Bonche}
\affiliation{Service de Physique Th\'eorique
             CEA-Saclay, 91191 Gif sur Yvette Cedex, France}
\date{August 12, 2004}
\begin{abstract}
The influence of exact angular-momentum projection and configuration
mixing on properties of a heavy, well-deformed nucleus is discussed 
for the example of \nuc{240}{Pu}. Starting from a self-consistent 
model using Skyrme interactions, we analyze the resulting modifications 
of the deformation energy, the fission barrier height, the excitation 
energy of the superdeformed minimum associated with the fission isomer, 
the structure of the lowest rotational bands with normal deformation 
and superdeformation, and the corresponding quadrupole moments and 
transition moments.
We present results obtained with the Skyrme interactions SLy4 and SLy6, 
which have slightly different surface tensions.
\end{abstract}
\pacs{21.60.Jz, 
      21.30.Fe, 
      21.10.-k, 
      27.90.+b  
}
\maketitle
%
%
\section{Introduction}
Microscopic mean-field methods~\cite{RMP} are particularly
well suited to describe nuclei with a well defined shape. When the energy
of a nucleus depends softly on a shape degree of freedom or presents
several minima  as a function of this shape,
correlations beyond mean field can affect
the properties of the ground state strongly. In such cases,
the two most relevant types of correlations are associated with the 
rotation of the nucleus and with its vibrations with respect to deformation.
The inclusion of rotational correlations can be performed by
a symmetry restoration and that of vibrations
by a mixing of mean-field states corresponding to different shapes.
In both cases,  this requires to go beyond  mean-field models.

We have recently developed a method which achieves these 
goals~\cite{Val00a}. Applications have been carried out for neutron-deficient 
Pb isotopes \cite{Dug03pb,Ben03pb}.
The low-energy spectrum of these nuclei varies rapidly with neutron number
with states exhibiting strong mixing between oblate, spherical and prolate 
configurations. Qualitative properties of their spectra, including transition 
properties, were nicely explained. However, since the results depend
strongly on the amount of mixing between several configurations,
a detailed agreement with the data has not been achieved.

A very different situation occurs when the mean-field approximation
is better justified as a first approximation, such as when coexisting states 
lie in well separated energy minima. This is the case at low excitation 
energies for superdeformed bands in nuclei around Hg and for fission 
isomers~\cite{Hee97a}.
Nevertheless, it is only using beyond mean-field models that one can calculate 
spectra with well defined spin assignments as well as the corresponding 
transition probabilities.

The nucleus \nuc{240}{Pu} has often been used as a benchmark to study 
mean-field theories and effective interactions. We present here an 
application of our method to this nucleus. It will allow us to address 
the following issues:

(i) are quadrupole correlations influencing a well-deformed nucleus
\emph{a priori} well-described by mean-field calculations?

(ii) how does the exact angular momentum projection modify the fission 
barrier and the excitation energy of fission isomers?

(iii) how much do the predictions depend upon different parameterizations
of the effective interaction?

In what follows, we shortly recall the basic ingredients of the theory,
then we present our results for the spectrum of \nuc{240}{Pu} 
in the ground state and the superdeformed well. The fission barrier 
obtained with two different effective interactions is discussed and 
compared with earlier, more phenomenological, approaches.
%
%
\section{The model}
The starting point of our method is a set of HF+BCS wave functions
generated by self-consistent mean-field calculations with a constraint 
on a collective coordinate, the axial quadrupole moment 
$q=\langle Q_{20} \rangle$ in the present study. In the language of 
the spherical nuclear shell model, such mean-field states incorporate 
particle-particle (pairing) correlations as well as many-particle 
many-hole correlations by allowing deformations of the nucleus in 
its intrinsic frame. As a consequence, the mean-field 
states break several symmetries of the exact many-body states.
This symmetry violation makes it difficult to relate 
mean-field results to spectroscopic data which are obtained
in the laboratory frame of reference.
The second step of our method is a restoration
of the symmetries associated with particle numbers and rotation.
Another ambiguity in the interpretation of mean-field results arises 
from the non-orthogonality of mean-field states 
corresponding to different quadrupole moments,
so that different minima in a potential landscape cannot always be safely 
associated with different physical states. 
This difficulty is resolved in the third step of our method
by variational mixing of symmetry-restored mean-field states 
corresponding to different quadrupole moments. The method that we use 
is a discretized version of the generator coordinate method.
It removes the contribution of vibrational excitations to
the ground state and, at the same time, permits to construct 
a spectrum of excited states. 

In our method, the same effective interaction is used to generate the 
mean-field states and to perform the configuration mixing. We present 
below results obtained with two different Skyrme interactions, SLy4 
and SLy6~\cite{SLyx}. In both cases  a density-dependent zero-range  
interaction is used in the pairing channel. We use the same strength
as in previous studies, $-1250$ MeV fm$^3$ and two cutoffs, above and 
below  the Fermi energy, as defined in Ref.~\cite{Rig99}. The two Skyrme 
parameterizations differ mostly by their surface tension: the surface 
energy coefficient obtained from 
Hartree-Fock calculations of semi-infinite nuclear matter is 
lower for SLy6, 17.74 MeV, than for SLy4, 18.37 MeV \cite{samynpriv}. 
Such a difference is expected to affect significantly the deformation 
energy at large quadrupole moments. Both parameterizations have been 
fitted in an identical way. Their differences have their origin in a 
different choice for the treatment of the spurious center-of-mass 
motion (c.m.): a fully variational c.m.\ one for SLy6 and a simpler 
one-body approximation for SLy4. The energy differences due to these 
two schemes induce slight differences in the properties of 
the interactions, see Ref.\ \cite{Ben00b} for a detailed discussion.

As our main goal is an investigation of the overall effect of 
symmetry restoration and configuration mixing in a heavy, well-deformed 
nucleus, we restrict ourselves to axial and reflection-symmetric 
shape degrees of freedom.
It has been shown that the fission barrier height 
obtained with SLy6 is in agreement with experiment within 1~MeV 
when octupole deformation is taken into account~\cite{RMP}.

Our method has many interesting properties.
Its sole phenomenological ingredient is the effective nucleon-nucleon 
interaction, which has been adjusted once and for all on generic 
nuclear properties. From a numerical point of view, it is simple 
enough to be applied  throughout the mass table up to superheavy 
nuclei, utilizing the full model space of single-particle states 
with the proper coupling to the continuum. Another attractive aspect 
of the method  is that it allows to determine electric transition 
probabilities directly in the laboratory frame between any pair of states.
Finally, the method has the advantage that its results can
be interpreted within the intuitive picture of intrinsic shapes
and shells of single-particle states that is offered by the framework
of mean-field models. More details on the method can be found in 
Refs.\ \cite{Val00a,Ben03pb}.

Spectroscopic quadrupole moments and $B(E_\lambda)$ values are 
determined directly in the laboratory frame of reference~\cite{BFH03}. 
To connect our results with other approaches, it is interesting to 
derive quantities analogous to intrinsic frame parameters from 
spectroscopic or transition moments. An intrinsic charge quadrupole 
moment $\Qzt$ can be determined from $B(E_2)$ values:
\be
\label{E02}
\Qzt
= \sqrt{\frac{16 \pi}{5}  
  \frac{B(E2,J \to J-2)}
       {\langle J \,0 \, 2 \, 0 | J-2 \, 0 \rangle^2 e^2}}
,
\ee
or can be related to the spectroscopic quadrupole moment 
$Q_c(J,k)$ via the relation
\be
\label{E04}
\Qzs 
= - \frac{2J+3}{J} \, Q_c(J,k)
.
\ee
We also adopt the sharp edge liquid drop relation to relate the $\bd$ 
deformation parameter and the axial quadrupole moment $Q_2$
\be
\label{E01}
\bd = \sqrt{\frac{5}{16 \pi}} 
\, \frac{4 \pi Q_2}{3 R^2 A}\quad,
\ee
where the nuclear radius $R$ in fm at zero deformation is related 
to  the mass $A$ according to the standard formula 
\mbox{$R = 1.2 \, A^{1/3}$}. 
%
%
\section{Results}
%
%
\begin{figure}[t!]
\centerline{\epsfig{file=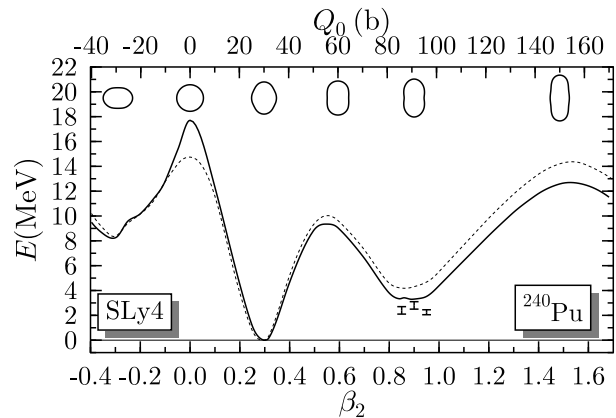}}
\caption{\label{fig:pu240:sly4:e}
Deformation energy curve of \nuc{240}{Pu} obtained with SLy4
projected on $N$ and $Z$ (dashed line) and projected 
on $N$, $Z$ and \mbox{$J=0$} (solid line). All energies are 
normalized to the deformed ground-state value of each curve.
The available experimental data for the excitation energy 
of the superdeformed band head are shown at arbitrary 
deformation (see text). Shapes along the path are indicated by the 
density contours at \protect\mbox{$\rho = 0.07$ fm$^{-3}$}.
}
\end{figure}
%
%
%
\subsection{Deformation energy}
There is a large set of data on fission barriers of actinide 
nuclei~\cite{Bjo80a,Spe74a}. Among them, the double-humped fission barrier 
of \nuc{240}{Pu} has been used as a benchmark for mean-field models 
and effective interactions. First calculations were performed with 
Skyrme forces and the Hartree-Fock+BCS method~\cite{Flo74}, with the
Gogny force and the HFB method~\cite{Gir83} or with relativistic  
Lagrangians and the relativistic mean-field method (RMF)~\cite{Blu94a}.
Semi-classical approximations of the mean-field method were
performed with different Skyrme forces in ref~\cite{Dut80a,ETFSI}.
Several Skyrme interactions and RMF Lagrangians were compared to the 
data in~\cite{Bue03a}. Axial and triaxial barriers obtained with 
Skyrme, Gogny and RMF forces are compared in Ref.\ \cite{RMP}.
Finally, the excitation energy of fission isomers has been
studied with a variety of Skyrme forces in Ref.\ \cite{Hee97a,Tak98a}.

\begin{figure}[t!]
\centerline{\epsfig{file=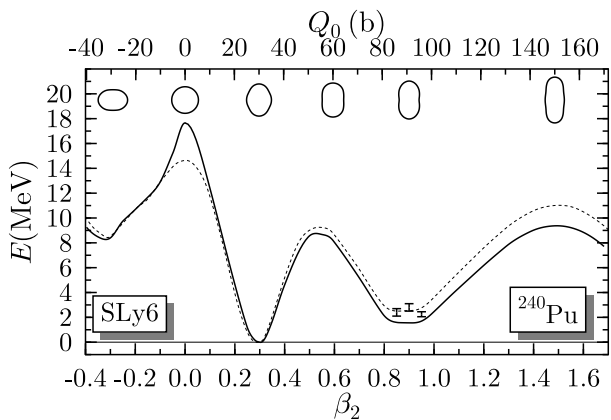}}
\caption{\label{fig:pu240:sly6:e}
Same as Figure \ref{fig:pu240:sly4:e}, for the interaction SLy6. 
}
\end{figure}

The deformation energy curves obtained after particle-number projection and 
particle-number + angular-momentum projection on $J=0$
are presented in Figures \ref{fig:pu240:sly4:e} and \ref{fig:pu240:sly6:e} 
for SLy4 and SLy6 interactions.
For all curves, the energy of the  ground state is taken as zero.
The ground state and  the fission isomer after projection are obtained 
from the mean-field minima. With the normalization that we have chosen
to plot the results, the gain of energy obtained for the ground state 
by angular momentum projection is given by the difference between 
the curves at spherical shape. This gain 
is around 3.0~MeV for both interactions,
bringing  the calculated total energy closer to the experimental one.
%
%
%
%

The fission isomer is obtained at a $\beta_2$ value around 0.9;
angular-momentum projection lowers its excitation energy by about 1~MeV
for both forces, from 4.3 to 3.3~MeV for SLy4, and from 2.6 to 1.6~MeV 
for SLy6. 
%
%
\subsection{Rotational energy}
\begin{figure}[t!]
\centerline{\epsfig{file=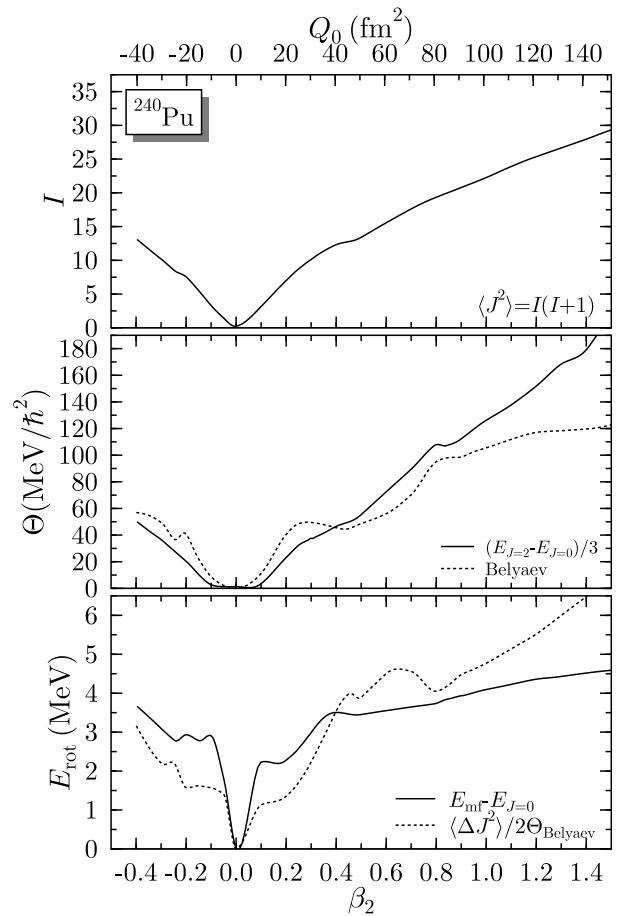}}
\caption{\label{fig:pu240:sly4:erot}
Top to bottom:
the average angular momentum $I$ of the mean-field states
obtained from $\langle J^2 \rangle = \hbar^2 \, I (I+1)$,
the moment of inertia calculated from the difference 
of the projected \mbox{$J=2$} and \mbox{$J=0$} energy curves 
(solid line) and the Belyaev moment of inertia (dotted line);
and the rotational energy obtained from the energy difference
between the mean-field and the \mbox{$J=0$} energy curves
(solid line) and the rotational correction, 
Eq.~(\ref{eq:rotcor}).
}
\end{figure}

Angular momentum projection provides the exact correction for
the spurious rotational energy of the mean-field states. It is given 
by the difference between the binding energies before and after projection
on angular momentum \mbox{$J=0$}
\begin{equation}
E_{\rm rot}(\beta_2)
= E_{\text{mf}}(\beta_2) - E_{J=0}(\beta_2)
.
\end{equation}
This difference does not depend much on the Skyrme parameterization and 
is plotted in the lower panel of Fig.~\ref{fig:pu240:sly4:erot} for SLy4. 
It is zero at spherical shape and
increases first rapidly to values around 3~MeV for deformations smaller 
than \mbox{$|\beta_2| < 0.1$}, and then moderately for larger deformations.
A similar behavior has been obtained in most of our previous 
calculations \cite{Ben03pb,BFH03}. The topology of the fission 
barrier is not much affected by angular momentum projection. The  
height of the second barrier is decreased by about 800~keV with 
respect to the fission isomer and by 1.5~MeV with respect to the 
ground state.
%
%
%
%
%
%
\begin{figure*}[t!]
\centerline{\epsfig{file=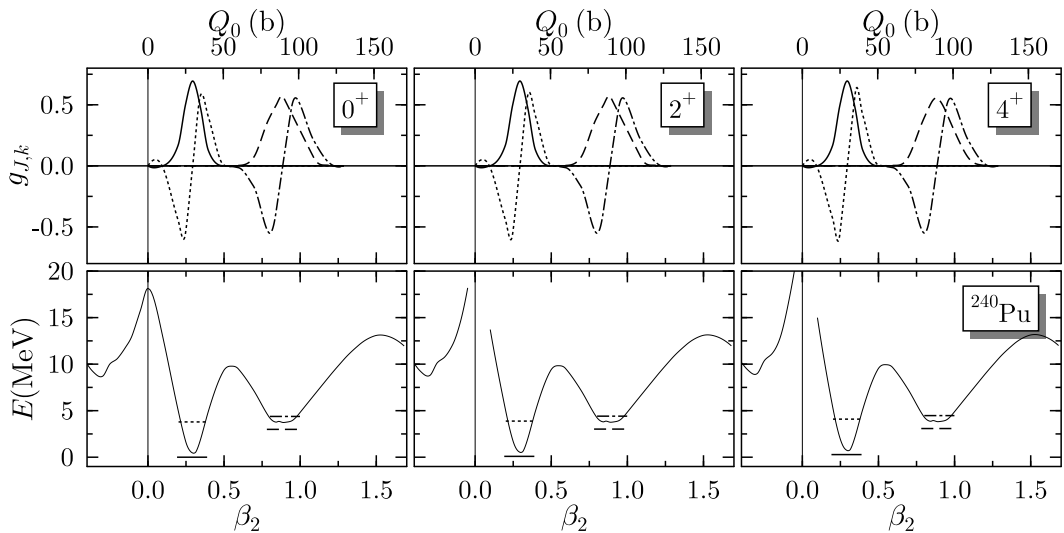}}
\caption{\label{fig:pu240:sly4:wf}
Collective wave functions (upper panel) of the lowest $0^+$, $2^+$, 
and $4^+$ states.
The lower panels display the corresponding 
excitation energies at the average deformation of the mean-field
states from which they are built, together with the projected energy
curve.}
\end{figure*}
%
%

Rotational corrections are sometimes incorporated phenomenologically 
as a perturbation to mean-field calculations. In particular, it is an
ingredient in the mass formulae based on Skyrme forces~\cite{Gor03}.
In the same way, the fission barrier of \nuc{240}{Pu} 
including a rotational correction was used as a constraint
in the fit of the Gogny interaction~\cite{Gir83}. In both cases, 
the rotational energy has the form
\begin{equation}
\label{eq:rotcor}
\tilde{E}_{\rm rot}(\beta_2)
=  \frac{\langle J^2 \rangle_{\beta_2}}
        {2\Theta(\beta_2)}
,
\end{equation}
where $\langle J^2 \rangle$ is the mean value of the square of the angular
momentum for the mean-field state, and the moment of inertia $\Theta$ is 
determined from an approximate cranking formula
\begin{equation}
\label{eq:belyaev}
\Theta_{\text{Belyaev}}
= 2 \sum_{i,j > 0} \frac{|(i|\hat{J}_y|j)|^2}
                        {E_i + E_j} (u_i v_j - v_i u_j)
.
\end{equation}
The sum in Eq.~(\ref{eq:belyaev})
runs over the single-particle states $|i)$ and $|j)$ in a given 
deformed mean-field state, and $E_i$ and $E_j$ are the corresponding 
quasi-particle energies. For the large-scale mass fits of Ref.~\cite{Gor03}, 
the actual moment of inertia taken is a mixture of Eq.~(\ref{eq:belyaev})
and of the rigid-body moment of inertia. More involved approximations 
for the moment of inertia have been developed, see e.g.\ \cite{RS80,Rob04} 
and references therein, but are rarely used. We can extract a moment of 
inertia from our projected mean-field calculations using the energy 
difference between the energy curves for \mbox{$J=0$} and \mbox{$J=2$} as
\begin{equation}
E_{J=I}(\beta_2) - E_{J=0}(\beta_2) 
= \frac{\hbar^2 I (I+1)}
       {2 \Theta(\beta_2)}
.
\end{equation}
In Fig.~\ref{fig:pu240:sly4:erot}, we compare the exact and approximate
rotational energies as a function of deformation. We also show 
the angular momentum dispersion of the mean-field wave functions and
the moment of inertia given by Eqn.~(\ref{eq:belyaev}) and obtained
from the $2^+$ excitation energy. Both of these moments of inertia
are rather close for deformations around that of
the fission isomer. For lower deformations, the Belyaev moment of inertia
is significantly larger than the ``exact'' one. For larger deformations, 
the Belyaev moment does not increase as rapidly as the one extracted 
from our calculation. The dispersion of the angular momentum of the 
mean-field wave function increases 
also with deformation in a way which is partly compensated by the increase
of the Belyaev moment of inertia. However, this compensation is not strong
enough and the Belyaev rotational energy correction overestimates
at large deformations the energy correction obtained by exact projection.
Note that this correction varies only by 1.5~MeV from the deformation 
corresponding to the ground state to the external fission barrier.
%
%
\subsection{Configuration mixing}
The properties of the four lowest states obtained from the configuration 
mixing calculation are given in Table \ref{tab:sly4:bands} for SLy4
and Table \ref{tab:sly6:bands} for SLy6. The corresponding collective wave 
functions are shown in Figure~\ref{fig:pu240:sly4:wf} for the SLy4 interaction.
The states separate nicely into four rotational bands, two located in the 
prolate normal-deformed (ND) minimum, and two in the superdeformed (SD) one. 
The wave functions are all confined within either the ND or the SD wells.
States obtained with SLy6 are similar, except for an overall shift in 
the excitation energies of the states located in the superdeformed well; 
the SD $0^+$ band-head has an excitation energy of 2.99~MeV with SLy4 and 
1.25~MeV with SLy6. The actual experimental value is in between 
the two, although there are some conflicting values given in the literature 
$2.4 \pm 0.3$~MeV \cite{Bjo80a}, $\approx 2.8$ MeV \cite{isomers}
and $2.25 \pm 0.20$~MeV \cite{Hun01}.
%
%
\begin{table}[t!]
\begin{tabular}{lccccccc}
\hline\noalign{\smallskip}
\multicolumn{1}{c}{state} &
\multicolumn{1}{c}{$E$} &
\multicolumn{1}{c}{$Q_s$} &
\multicolumn{1}{c}{$Q_0^{(s)}$} &
\multicolumn{1}{c}{$\beta_2^{(s)}$} &
\multicolumn{1}{c}{$B(E2)\!\!\downarrow$} &
\multicolumn{1}{c}{$Q_0^{(t)}$} &
\multicolumn{1}{c}{$\beta_2^{(t)}$} \\
 & \multicolumn{1}{c}{(MeV)} &
\multicolumn{1}{c}{($e$\,b)} &
\multicolumn{1}{c}{($e$\,b)} & &
\multicolumn{1}{c}{($e^2$\,b$^2$)} &
\multicolumn{1}{c}{($e$\,b)} & \\
\noalign{\smallskip}\hline\noalign{\smallskip}
$ 0^+_1$ &   0.000 &     --   &    --    &    --   &         --   &    --    &    --    \\
$ 2^+_1$ &   0.083 &   -3.4   &   11.9   &   0.300 &      2.80    &   11.9   &    0.300 \\
$ 4^+_1$ &   0.277 &   -4.3   &   11.9   &   0.300 &      4.00    &   11.9   &    0.300 \\
\noalign{\smallskip}\hline\noalign{\smallskip}
$ 0^+_3$ &   3.793 &     --   &    --    &    --   &         --   &    --    &    --    \\
$ 2^+_3$ &   3.880 &   -3.4   &   11.9   &   0.301 &      2.82    &   11.9   &    0.301 \\
$ 4^+_3$ &   4.088 &   -4.3   &   12.0   &   0.303 &      4.07    &   12.0   &    0.302 \\
\noalign{\smallskip}\hline\noalign{\smallskip}
$ 0^+_2$ &   2.953 &     --   &    --    &    --   &         --   &    --    &     --   \\
$ 2^+_2$ &   2.978 &  -10.3   &   36.0   &   0.911 &     25.8     &   36.0   &    0.911 \\
$ 4^+_2$ &   3.045 &  -13.1   &   36.0   &   0.911 &     36.8     &   36.0   &    0.911 \\
\noalign{\smallskip}\hline\noalign{\smallskip}
$ 0^+_4$ &   4.338 &     --   &    --    &     --  &        --    &    --    &     --   \\
$ 2^+_4$ &   4.364 &  -10.4   &   36.5   &   0.922 &     26.4     &   36.5   &    0.922 \\
$ 4^+_4$ &   4.429 &  -13.3   &   36.5   &   0.922 &     37.8     &   36.5   &    0.922 \\
\noalign{\smallskip}\hline
\end{tabular}
\caption{\label{tab:sly4:bands}
Properties of the rotational bands of \nuc{240}{Pu} obtained with SLy4:
excitation energy $E$, spectroscopic quadrupole moment $Q_s$, corresponding
quadrupole moment $Q_0^{(s)}$ and dimensionless deformation $\beta_2^{(s)}$
in the intrinsic frame, reduced $E2$ transition probability 
$B(E2)\!\!\downarrow$, and corresponding quadrupole moment $Q_0^{(t)}$ 
and dimensionless deformation $\beta_2^{(t)}$ in the intrinsic frame.
}
\end{table}
%
%

The band head of the second ND band is obtained at 3.79~MeV for
both forces. Similarly, the excitation energy of the second SD 
band with respect to the first SD band head is 1.39~MeV for SLy4 
and 1.27~MeV for SLy6. This suggests that the excitation energies within
a well are fairly independent from the surface tension of the Skyrme force. 
Both interactions give also similar values for
the excitation energies within the bands.

The ground state band is known up to very high spin 
\cite{Hac98,Wie99,Janpriv}. It has been suggested that static octupole 
deformation plays a role to explain the behaviour at large angular momentum
Ref.\ \cite{Wie99,She00}. 
States below the $6^+$ decay mainly by internal electron conversion, 
so the corresponding transitions have not been detected in $\gamma$-ray 
spectroscopy. 
%
%
\begin{table}[t!]
\begin{tabular}{lccccccc}
\hline\noalign{\smallskip}
\multicolumn{1}{c}{state} &
\multicolumn{1}{c}{$E$} &
\multicolumn{1}{c}{$Q_s$} &
\multicolumn{1}{c}{$Q_0^{(s)}$} &
\multicolumn{1}{c}{$\beta_2^{(s)}$} &
\multicolumn{1}{c}{$B(E2)\!\!\downarrow$} &
\multicolumn{1}{c}{$Q_0^{(t)}$} &
\multicolumn{1}{c}{$\beta_2^{(t)}$} \\
 & \multicolumn{1}{c}{(MeV)} &
\multicolumn{1}{c}{($e$ b)} &
\multicolumn{1}{c}{($e$ b)} & &
\multicolumn{1}{c}{($e^2$ b$^2$)} &
\multicolumn{1}{c}{($e$ b)} & \\
\noalign{\smallskip}\hline\noalign{\smallskip}
$ 0^+_1$ &   0.000 &    --    &     --   &    --   &         --   &     --   &     --   \\
$ 2^+_1$ &   0.083 &   -3.4   &   11.9   &   0.300 &      2.81    &   11.9   &    0.300 \\
$ 4^+_1$ &   0.273 &   -4.3   &   11.9   &   0.301 &      4.02    &   11.9   &    0.301 \\
\noalign{\smallskip}\hline\noalign{\smallskip}
$ 0^+_4$ &   3.794 &     --   &     --   &    --   &         --   &     --   &     --   \\
$ 2^+_4$ &   3.879 &   -3.4   &   12.0   &   0.304 &      2.88    &   12.0   &    0.304 \\
$ 4^+_4$ &   4.082 &   -4.4   &   12.1   &   0.306 &      4.15    &   12.1   &    0.305 \\
\noalign{\smallskip}\hline\noalign{\smallskip}
$ 0^+_2$ &   1.251 &     --   &     --   &    --   &         --   &     --   &     --   \\
$ 2^+_2$ &   1.277 &  -10.4   &   36.3   &   0.916 &     26.1     &   36.2   &    0.915 \\
$ 4^+_2$ &   1.338 &  -13.2   &   36.3   &   0.917 &     37.4     &   36.3   &    0.916 \\
\noalign{\smallskip}\hline\noalign{\smallskip}
$ 0^+_3$ &   2.519 &     --   &     --   &    --   &         --   &     --   &     --   \\
$ 2^+_3$ &   2.550 &  -10.5   &   36.7   &   0.928 &     27.0     &   36.8   &    0.931 \\
$ 4^+_3$ &   2.611 &  -13.4   &   36.7   &   0.928 &     38.3     &   36.7   &    0.928 \\
\noalign{\smallskip}\hline
\end{tabular}
\caption{\label{tab:sly6:bands}
The same as Table~\ref{tab:sly4:bands}, but for SLy6.
}
\end{table}
%
%

The lowest levels in the ground-state band are reported in the NUDAT
data base~\cite{nudat} at 42.824 ($2^+$), 141.690 ($4^+$), and 
294.319~keV ($6^+$). Our calculation overestimates these energies by 
almost a factor two. The experimental energies in the SD well are
20.1~keV for the $2^+$ and 66.8 keV for the $4^+$ for the 
SD1 band, and 769.9 ($0^+$), 785.1 ($2^+$), and 825.0~keV 
($4^+$) for the SD2 band. They are also overestimated by our model. 
On the contrary, the calculated $B(E2; 2^+ \to 0^+)$ 
value is in excellent agreement with the experimental one
of $26660 \pm 360$ $e^2$~fm$^4$ obtained from Coulomb excitation 
\cite{Bem73aE}.

A similar overestimation of excitation energies has been found for 
other nuclei with our model \cite{Val00a,Ben03pb} and in a 
similar framework using a Gogny interaction \cite{Rod00a}.
A hint on a possible origin of this discrepancy can be found from
cranked mean-field calculations. In this case, neither configuration 
mixing nor restoration of symmetries are performed. However, time-reversal 
invariance is broken and the mean-field potential is optimised for each $J$ 
value and not only for \mbox{$J=0$}. An unprojected cranked HFB calculation 
\cite{Ben03c} using the same effective interaction as here gives excitation 
energies for the ground-state band of  0.030 ($2^+$), 0.121 ($4^+$) and 
0.271~MeV ($6^+$), respectively, in much better agreement with the 
data. Looking at Figure \ref{fig:pu240:sly4:wf} one can see that, within 
a band, the amplitudes of the collective wave functions for different 
$J$ values differ by less than 1$\%$ for each deformation. 
The wave functions for different $J$ values are therefore probably too close 
to each other in our model. The slight breaking of time-reversal symmetry of
mean-field states subject to a cranking constraint might be sufficient to
provide a better starting point for exact projection and configuration mixing
for the $J$ different from 0 states.

The deformation of the ground state stays remarkably constant at all
levels of approximations: from \mbox{$\beta_2 = 0.29$} for the minima of
the mean-field  and projected energy curves to 
\mbox{$\beta_2^{(s)}(2^+) = 0.30$} as deduced from Eqn.\ (\ref{E01}) 
for both the spectroscopic and the transition quadrupole moments.
All these deformations agree with the one deduced from the 
experimental $B(E2)$ value, \mbox{$\beta_2 = 0.29$}. 

Since we obtain nearly equal $\beta_2^{(s)}$ and $\beta_2^{(t)}$ values
all along the bands, the use of the rotor model is well justified to
describe the four bands. 

We obtain also very large $E0$ transition matrix elements between states
in the same well. With SLy4, the $M(E0; 0^+_3 \to 0^+_1)$ in the first
well has a value of -29 $e$ fm$^2$ corresponding 
to $\rho(E0) = -0.52$, while the the $M(E0; 0^+_4 \to 0^+_2)$ in the
second well is 87 $e$ fm$^2$, or $\rho(E0) = 1.6$. As our
model predicts the collective wave functions of all members of a rotational
band to be very similar, the transition moments are calculated to be
very similar as well for all $E0$ transitions within a band. The values
obtained with the SLy6 interation differ only marginally from those
of SLy4. The $E0$ matrix elements that we obtain for transitions 
between states in the SD and the normal-deformed well are very small.
%
%
\section{Discussion and summary}
The inclusion of correlations beyond mean-field confirms that many properties
of the \nuc{240}{Pu} nucleus are already well described at the mean-field level 
of approximation.
The overall structure of the potential energy curve is not altered by angular
momentum projection, with a well defined prolate ground state and a fission
isomer in  narrow potential wells. Thanks to that, 
the properties of the lowest state in each well after
configuration mixing are very close to the properties
of the mean-field minima.
The ground-state and fission isomer wave functions have a Gaussian
shape and do not spread much around their respective minima.
The superdeformed minimum in the potential energy surface is, 
however, too wide to confine completely the wave functions of the lowest 
SD band which are more spread, but still of Gaussian shape.

As expected, the total binding energy is slightly increased by 
angular momentum projection, and thereby comes closer to the
experimental value. Compared to the ground state, angular-momentum 
projection lowers the (axial) inner barrier by about 0.6~MeV, the 
fission isomer by about~1 MeV, and the (reflection-symmetric) 
outer barrier by about 2~MeV. 
These changes of the potential landscape are going in the same direction 
as a decrease of the surface tension of the effective mean-field interaction 
and should therefore be considered when predicting superdeformed band heads 
and fission barrier heights. The schematic rotational correction used 
in the literature gives a too large reduction of the fission barrier 
heights by at least 2 MeV.

The energy of the
superdeformed fission isomer turns out to be too low with SLy6, 
while for SLy4 it is slightly too high. We obtained similar results 
for SD band heads of Pb isotopes in the \mbox{$A \approx 190$}
region \cite{Hee97a}. This suggests, that the surface tension might be 
too low for SLy6, while it is slightly too high for SLy4. This is 
apparently in contradiction with Ref.\ \cite{Ben00b}, where it was 
argued on the basis of pure mean-field calculations for \nuc{240}{Pu}, 
that the surface tension of SLy6 is more realistic than that of SLy4.
However, surface tension is not the only ingredient responsible for
the energy of superdeformed states, as their existence is usually 
caused by a shell effect. Unfortunately, it is
hard to disentangle the contribution from 
the macroscopic properties of the forces from the shell structure 
which is also not identical for SLy4 and SLy6 at large deformations.

The description of excited states in both wells is only
partly satisfactory. While we obtain quite good results for $E2$ 
transition moments and deformations, neither the excitation energies 
of the excited $0^+$ band heads nor the excitation energies within 
the bands are well reproduced. Unprojected cranked HFB calculations 
without any additional correlations perform much better for in-band 
transitions. This strongly suggests to extend our method to the use 
of such states as a starting set of wave functions. Such a development 
is currently underway. Using cranked mean-field states, however, will 
not change the too large excitation energies of $0^+$ band heads.
%
%
\begin{acknowledgments}
This work was supported by the PAI-P5-07 of the PPS
Scientific Policy. MB acknowledges support from the European 
Community. MB and PHH thank for the hospitality at the Institute
for Nuclear Theory, Seattle, where this work was finalised.
\end{acknowledgments}
%
%

\end{document}